\documentclass[notitlepage,twocolumn,prl,tightenlines,nofootinbib,superscriptaddress,showpacs]{revtex4}

\usepackage{amsmath}
\usepackage{amssymb,amsfonts,latexsym}
\usepackage{bm}
\usepackage[mathcal]{euscript}
\usepackage{graphicx}
\usepackage{epsfig}
\usepackage{color}

\definecolor{blue}{rgb}{0,0,1}
\definecolor{green}{rgb}{0,0.5,0}
\definecolor{red}{rgb}{1,0,0}
\definecolor{pink}{rgb}{0.9,0.3,0.7}
\definecolor{azur}{rgb}{0,0.5,0.5}
\definecolor{orange}{rgb}{1,0.5,0.2}
\definecolor{brown}{rgb}{0.5,0,0}

\newcommand{\be}{\begin{equation}}
\newcommand{\ee}{\end{equation}}
\newcommand{\ben}{\begin{equation*}}
\newcommand{\een}{\end{equation*}}
\newcommand{\ba}{\begin{eqnarray}}
\newcommand{\ea}{\end{eqnarray}}

\begin{document}

\title{Avalanches and Dynamical Correlations in supercooled liquids}
\author{R. Candelier}
\affiliation{SPEC, CEA-Saclay, URA 2464 CNRS, 91 191 Gif-sur-Yvette, France}
\author{A. Widmer-Cooper}
\affiliation{Materials Sciences Division, Lawrence Berkeley National Laboratory, Berkeley, California, 94720, USA}
\author{J. K. Kummerfeld}
\affiliation{Department of Chemistry, University of Sydney, Sydney, New South Wales 2006, Australia}
\author{O. Dauchot}
\affiliation{SPEC, CEA-Saclay, URA 2464 CNRS, 91 191 Gif-sur-Yvette, France}
\author{G. Biroli}
\affiliation{Institut de Physique Th\'eorique, CEA, IPhT, F-91191 Gif-sur-Yvette, France and CNRS, URA 2306}
\author{P. Harrowell}
\affiliation{Department of Chemistry, University of Sydney, Sydney, New South Wales 2006, Australia}
\author{D.R. Reichman}
\affiliation{Columbia University, 3000 Broadway, New York, New York, 10027, USA}

\begin{abstract}
We identify the pattern of microscopic dynamical relaxation for a two
dimensional glass forming liquid.  On short timescales, bursts of
irreversible particle motion, called cage jumps, aggregate into
clusters.  On larger time scales, clusters aggregate both spatially
and temporally into avalanches.  This propagation of mobility, or
dynamic facilitation, takes place along the soft regions of the
systems, which have been identified by computing isoconfigurational
Debye-Waller maps. Our results characterize the way in which dynamical
heterogeneity evolves in moderately supercooled liquids and reveal that
it is astonishingly similar to the one found for dense glassy granular
media. 
\end{abstract}

\maketitle
\vspace{-2mm}
Identifying the physical mechanisms responsible for the slowing down
of the dynamics of supercooled liquids is still an open problem
despite several decades of intense research. While traditional
descriptions of glassy systems have mainly focused on energy landscape
concepts\cite{Goldstein} and spatially averaged quantities, recent
work has centered on the real space properties reflected in the
dramatically heterogeneous dynamics shared by nearly all glass-forming
materials.  Concomitantly, investigations of the behavior of dense
driven granular media have uncovered tantalizing similarities with the
dynamics of supercooled liquids\cite{marty2005sac,abate2006aja,ohern2002rpf,liu1998jnc}
and provided new inspirations for research on the glass transition. 
Among the most notable findings related to the real space dynamical
properties in supercooled liquids as well as granular media is the evidence that
dynamic facilitation~\cite{butler1991,garrahan2002prl} and sizable
dynamic correlations~\cite{berthier2005dee} play an important
role. This is supported by the detailed analysis of microscopic
dynamics which has identified correlated particle motion in clusters,
strings and other motifs.~\cite{donati1999scm,candelier2009bbo,candelier2009eod}.
A natural question~\cite{ediger1996sla} related to these findings is
what, if any, structural features are correlated with the
heterogeneity noted in the real space dynamics.  Important progress in
this direction has been obtained~\cite{harrowell-propensity}, through
the introduction of the quantitative notion of ``propensity'', and
then later in~\cite{harrowell2008irs,brito2007hdm,yamamoto2009}, where it has been
shown that irreversible motion is correlated with the spatial
characteristics of soft modes.  

In this paper we address the following three questions:
First, do supercooled liquids exhibit the same hierarchical organization
of dynamics (i.e. cage escapes within clusters within avalanches) as
recently reported in granular materials~\cite{candelier2009bbo,candelier2009eod}?
Second, to what degree are these different scales of collective motion determined
by the underlying structure? And reciprocally, to what extend is the evolution of
the structure related to the relaxation events within a given realization of the dynamics?

We shall address these questions by performing computer
simulations on a new two-dimensional model of glass-forming liquid and
applying the cluster analysis developed in~\cite{candelier2009bbo}.
This new model is distinguished from previous 2D mixtures~\cite{harrowell}
in that supercooled liquid dynamics may be simulated without the formation
of palpable crystalline micro-domains.
Our main results are that the glassy dynamics of dense driven granular
systems~\cite{candelier2009bbo} and supercooled liquids turn out to be
astonishingly similar even at the microscopic level. {\em This is
remarkable given the fact that granular systems are driven
non-equilibrium systems with dissipative contact interactions while
supercooled liquids are equilibrium conservative systems.}  Quasi-instantaneous
clusters of nearby relaxing particles are typically followed by
adjacent clusters showing how long term dynamical correlations emerge.
This dynamic facilitation leads to the formation of finite size and finite
duration avalanches located on the ``soft'' regions of the configuration
as probed by the iso-configurational average of the Debye-Waller factor. Finally
the clusters of relaxing particles induce non-local reorganisation of
the structure as probed by the dynamics itself of the Debye-Waller factor.

As a model for a supercooled liquid we focus on a $2D$ non-additive
binary mixture of $N = 5,760$ particles enclosed in a square box with
periodic boundary conditions, interacting via purely repulsive
potentials of the form $u_{ab}(r)=\varepsilon(\sigma_{ab}/r)^{12}$.
The mole fraction of the smaller particles is taken to be
$x_{1}=0.3167$. All units are reduced so that
$\sigma_{11} =\varepsilon = m = 1.0$, $m$ being the mass of both types of
particle. We use non-additive potentials, namely
$\sigma_{12} = 1.1\times \sigma_{11}$ and $\sigma_{22} = 1.4 \times \sigma_{11}$ to
avoid the formation of crystalline domains. The temperature dependence
of the structural and dynamical properties of this model were characterized 
in~\cite{widmer-cooper-phd}. Molecular dynamics
simulations were carried out at constant NVT (T=0.4) using the
Nose-Poincare Hamiltonian~\cite{bond1999} after equilibration at
constant NPT as described in~\cite{harrowell2008irs}. All time
units are scaled in such a way that the structural relaxation time $\tau_\alpha$,
defined as the time required for the self intermediate scattering function to decay of
$1/2$, equals $10^3$. The typical collision time is $0.12$ in
these units.

We choose as a measure of the local mobility (or relaxation) of a particle $p$:
\begin{equation}
Q_{p,t}(a,\tau) = \exp \left( - \frac{||\Delta \vec{r}_p(t,t+\tau)||^2}{2a^2} \right),
\label{def_de_Q}
\end{equation}
\noindent
where $\Delta \vec{r}_p(t,t+\tau)$ is the displacement of the particle
$p$ between $t$ and $t+\tau$ and $a$ is the length scale over which
the motion is probed. A global measure of the dynamics is provided by
the correlation function, $Q_t(a,\tau)=\frac{1}{N} \sum_p Q_{p,t}(a,\tau)$,
and its fluctuations $\chi_4(a,\tau)=N {\mathrm Var}\left(Q_t(a,\tau)\right)$.
As in~\cite{lechenault2008csa}, we focus on the values of $a$ and $\tau$ corresponding
to maximal dynamic heterogeneity, i.e. highest value of $\chi_4(a,\tau)$
(see~\cite{lechenault2008csa} for details). This leads to $a^*=0.29$ and
$\tau^*=1078$.  Note that the latter is very close to the relaxation time $\tau_\alpha=1000$.

\begin{figure}[t!] 
\center
\vspace{-0.5cm}
\includegraphics[width=0.49\columnwidth]{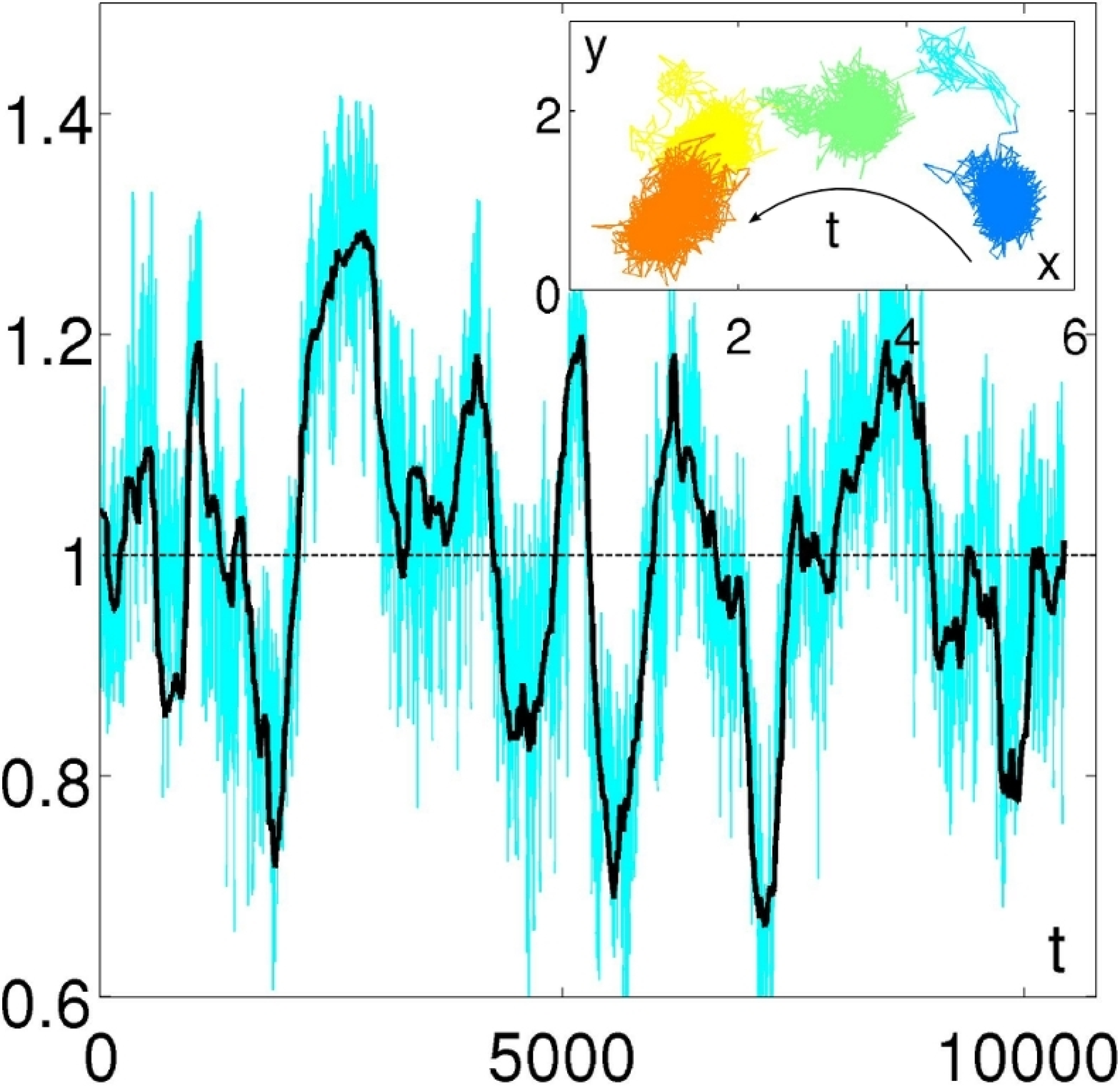}
\includegraphics[width=0.49\columnwidth]{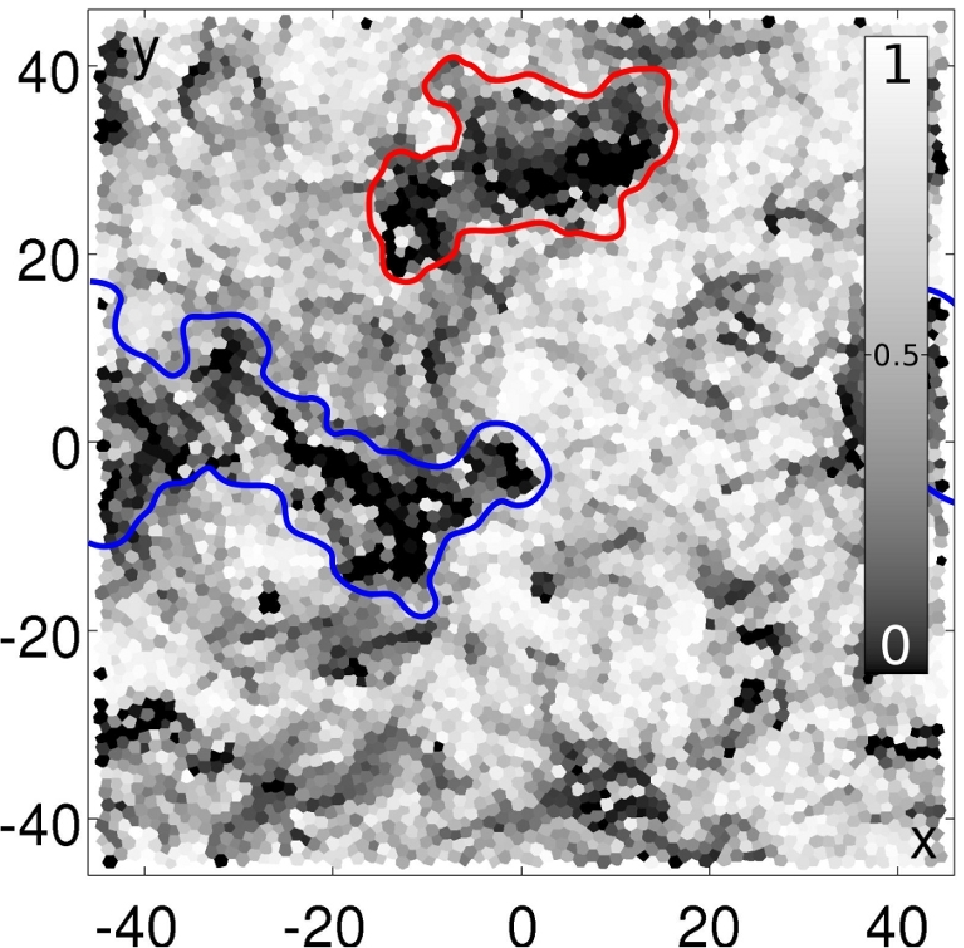}
\includegraphics[width=0.49\columnwidth]{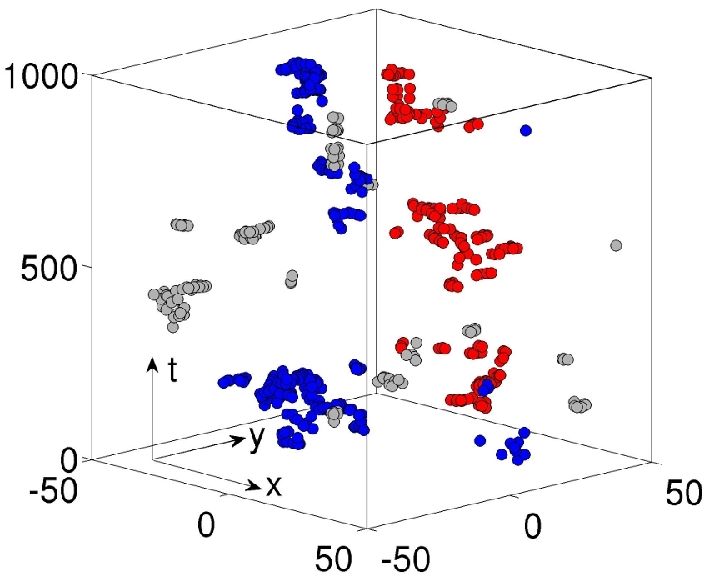}
\includegraphics[width=0.49\columnwidth]{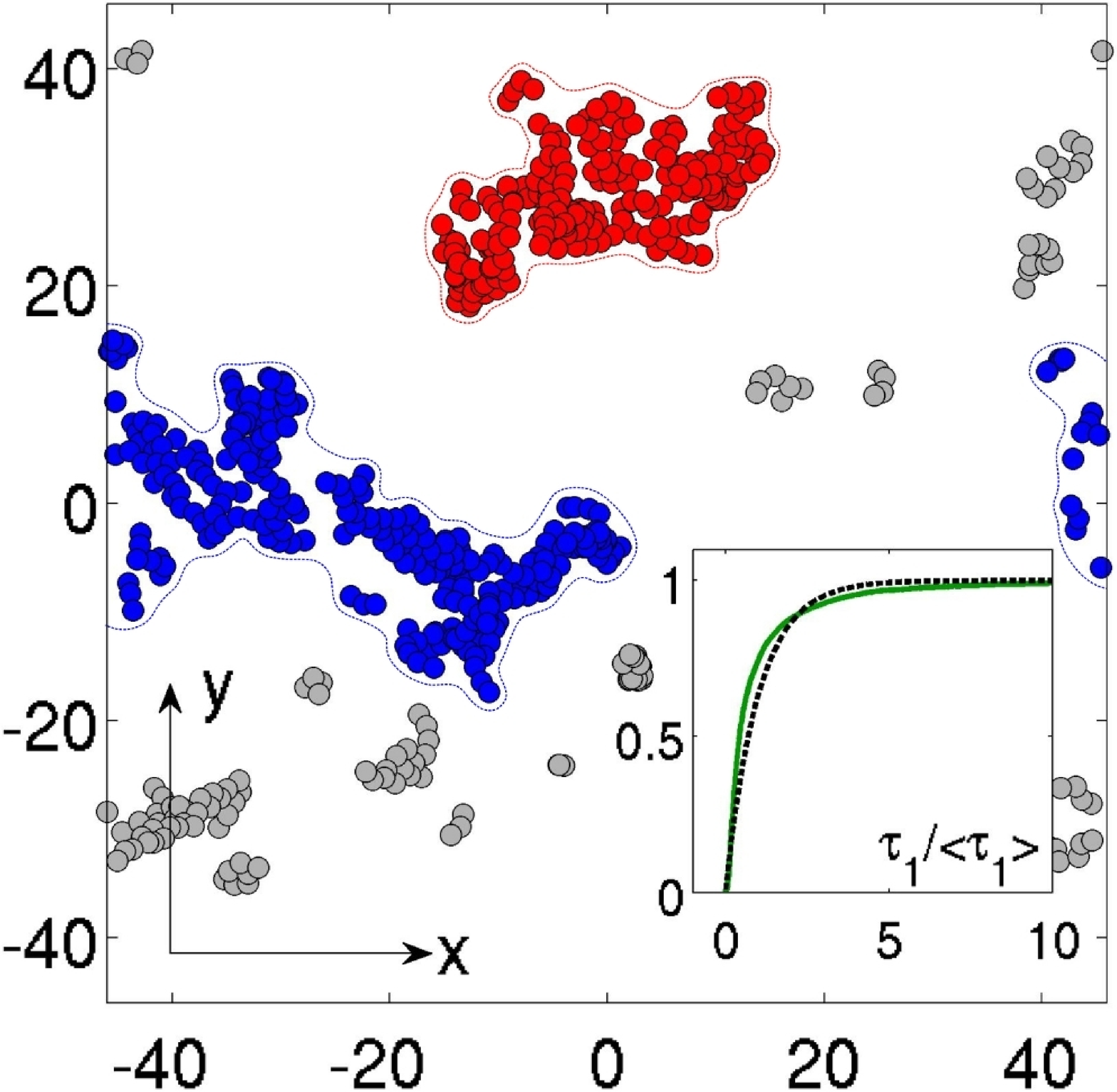}
\caption{(color online) Cooperative cage jumps form large decorrelation patterns.  Top left: Comparison between the relative averaged relaxation $Q_t(a^*,\tau^*)/\langle Q_t\rangle_t$ (cyan) and the relative percentage $P_t(\tau*)/\langle P_t\rangle_t$ of particles that have not jumped between $t_0$ and $t_0+\tau^*$ (black).  \textit{Inset} Trajectory of a single particle over $14 \tau_{\alpha}$. Color changes when the particle jumps.  Top right: Map of $Q_{t}(\tau^*)$.  Bottom left: Spatio-temporal view of the cage jumps between $t$ and $t+\tau^*$. The jumps corresponding to two arbitrarily chosen avalanches are painted in blue and red.  Bottom right: Map of the cage jumps occurring between $t$ and $t+\tau^*$. Note that the blue and red avalanches lie in distinct regions of space.  \textit{Inset} Cumulative Pdf of the reduced time lag between adjacent clusters $\tau_1$ (green), compared to the corresponding curve for a random distribution of clusters (black).}
\label{fig:CJ}
\vspace{-0.5cm}
\end{figure}

In order to analyze the microscopic dynamics and study possible
connections with the dynamics of dense driven granular media, we
follow the same procedure as in~\cite{candelier2009bbo}. This allows
one to separate the dynamics along a given trajectory into periods of
inefficient vibrational motion separated by relaxation events
also called cage jumps. (see inset of the top left panel of fig.~\ref{fig:CJ}). 
One has to bear in mind that a particle undergoing a cage jump does not
necessarily change neighbors.
In the top left panel of fig.~\ref{fig:CJ} we compare the relative
values of $Q_t(a^*,\tau^*)$ to those of $P_t(\tau^*)$, which is the
percentage of particles that have not jumped during the time $\tau^*$.
The two curves track each other, showing that cage jumps provides a
powerful coarse grained description of the dynamics. 
In addition, we also show that the cage jumps are exactly located in the areas where
the decorrelation is maximal (compare fig.~\ref{fig:CJ} right top and
bottom). We repeat the same spatio-temporal analysis performed for the
two dimensional granular media studied in~\cite{candelier2009bbo}. 
The outcome is remarkably similar. First, cage jumps aggregate into
clusters, which are formed by cage jumps adjacent in space (as
measured by the neighboring particles) and time (separated by less
than $\tau_{th}=28$, which is twice the precision of the cage
detection algorithm). The size of these clusters are largely
distributed with an average value of $7.6$ cage jumps
per cluster. Second, clusters aggregate into avalanches
in which the first cluster triggers the appearance of successive
clusters nearby shortly after, see fig.~\ref{fig:CJ} bottom left. 
This is clearly demonstrated as in~\cite{candelier2009bbo} by
focusing on the cumulative $Pdf$ of the lag times $\tau_1$
separating each cluster from the nearest adjacent one,
normalized by its average value $\langle \tau_1 \rangle$. See the
inset of fig.~\ref{fig:CJ},bottom right, where this $Pdf$ is compared
to the equivalent distribution for randomly distributed clusters in
space and time. One can see a clear excess of both small and large lags: 
the $Pdf$ can be extremely well fit by the union of
two data sets corresponding to Poissonian processes with two different
timescales $\tau_S=240$ and, $\tau_L=1746$. The short time scale
corresponds to the existence of a correlation among
adjacent clusters. The large one is related to the average time spent
in a cage. This leads to a very peculiar type of dynamical correlation,
which in the literature is often called dynamical facilitation~\cite{butler1991,garrahan2002prl}:
local relaxations are followed closely in space and in time by other local relaxations. The concatenation
of these events leads to the dynamical heterogeneity observed on the timescale $\tau_\alpha$.  
However, at least for the degree of supercooling considered here, we find that facilitation is not conserved in the following sense: avalanches are well separated, indicating that there are relaxation events, which cannot be explained by the facilitation mechanism.

\begin{table}[b!]
\vspace{-5mm}
\begin{tabular}{|r|c|c|c|c|c|c|}
\hline
& $a^*$ & $\xi_4$ & $\tau_\alpha$ & $\tau^*$ & $\tau_S$ & $\tau_L$ \\
\hline
Supercooled Liquid & $0.29$ & $2.9$ & $1000$ & $1078$ & $240$ & $1746$\\ \hline
Dense Granular Media & $0.12$ & $3.1$ & $1000$ & $915$ & $155$ & $1384$\\ \hline
\end{tabular}
\caption{Comparison of length and time scales normalized so that $\tau_\alpha=1000$. See definitions in the text.}
\label{tab:comp}
\end{table}

It is interesting to compare the actual values of these parameters to
those of the granular system investigated previously. This comparison
is performed in Table I, where we also report the value of the
dynamical correlation length $\xi_4$, obtained from the spatial range
of the dynamical correlator $G_4$(whose integral is equal to
$\chi_4$), see e.g.~\cite{glotzer2000tdf}. The dynamics are strikingly
similar, a non-trivial result given the difference between an equilibrated
thermal liquid and a non-equilibrioum steady state of vibrated grains.
One  difference we find is that the average distance between avalanches
is somewhat smaller in the liquid case than in the granular one:
$\sim10$ as compared to $\sim27$.
Recent results~\cite{candelier2009eod} obtained by changing the
density of the granular sample show that our model of a supercooled
liquid would compare with a granular system characterized by a slightly smaller density.

\begin{figure}[t!] 
\center
\vspace{-0.5cm}
\includegraphics[width=0.49\columnwidth]{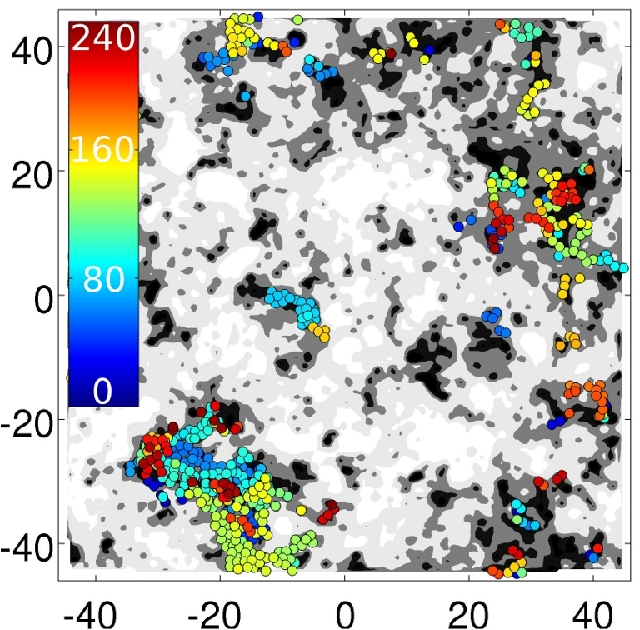}
\includegraphics[width=0.49\columnwidth]{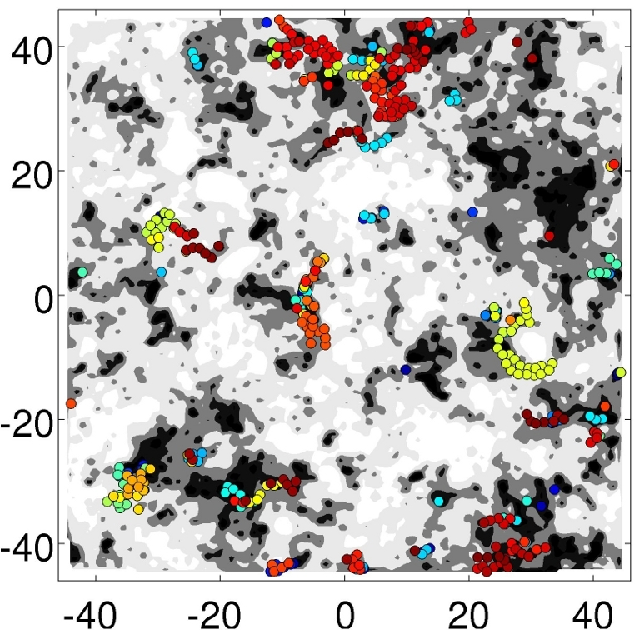}
\includegraphics[width=0.49\columnwidth]{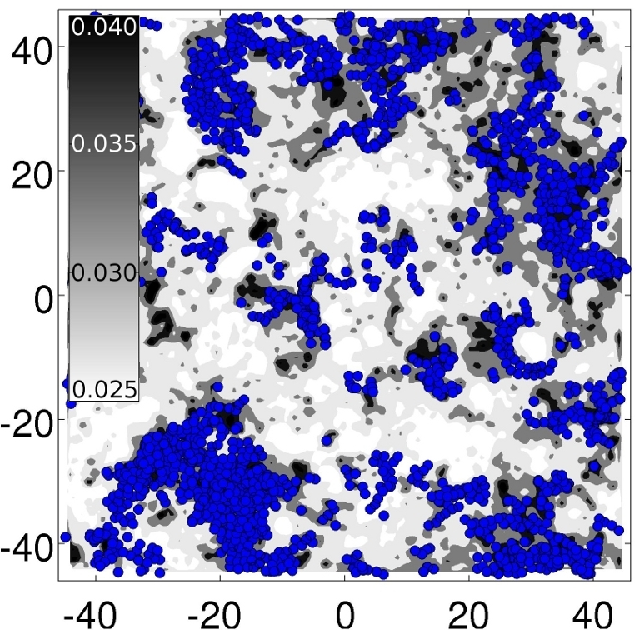}
\includegraphics[width=0.49\columnwidth]{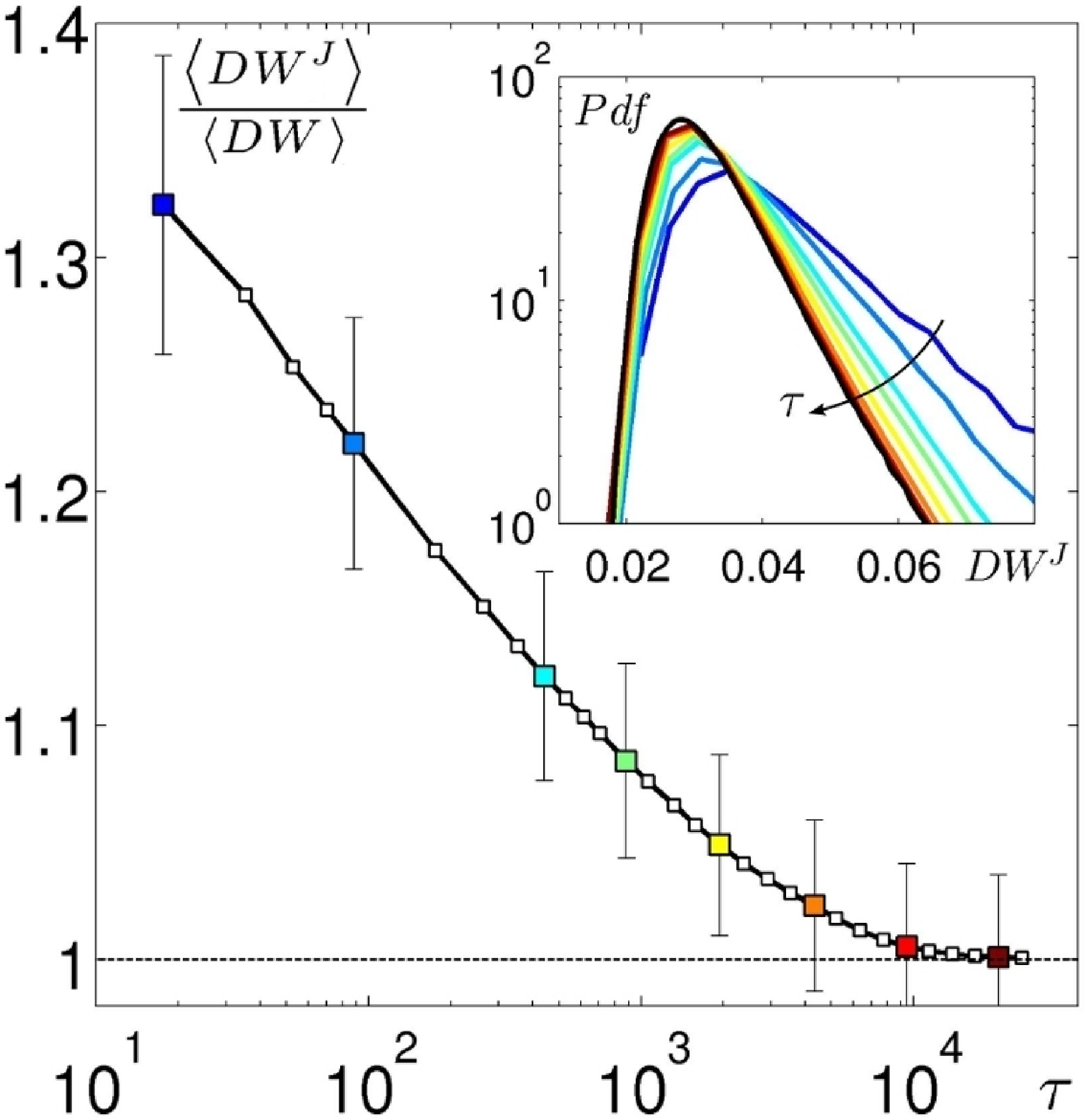}
\caption{(color online) Top: Cage jumps occuring between $t$ (blue) and $t + \tau_S$ (red) for two different isoconfigurational trajectories, on top of a $DW$ factor map computed at time $t$ (in grey). Bottom left: Cage jumps occurring in $6$ isoconfigurational trajectories between $t$ and $t+\tau_S$ (in blue) tile the high $DW$ regions. Colorbar indicates the $DW$ values in levels of grey.  Bottom right: Average $\langle DW^J\rangle$ over the particles having jumped between $t$ and $t+\tau$, divided by the average $\langle DW\rangle$ over all particles, as a function of the lag time $\tau$.
\textit{Inset} Pdf of $DW^J$ for the particles jumping in $[t;t+\tau]$ for several values of $\tau$. The black curve is the Pdf for all particles.}
\label{fig:DW}
\vspace{-0.5cm}
\end{figure}

We shall now investigate whether one can
find a property closely connected to the structure at time $t$, which
would allow one to predict where clusters will appear in the future
and even some aspects of avalanche evolution. On the basis of previous
work~\cite{harrowell2008irs,brito2007hdm} a natural candidate for such
a feature are the so-called soft modes. 
Here we will use another means of identifying the location of "soft''
regions or modes by using the isoconfigurational
Debye-Waller (DW) factor\cite{harrowell2008irs,harrowellDW}.  Starting
from the system configuration at time $t$, one computes the local
Debye-Waller factor for particle $i$: 
$DW_i = \langle[\vec{r_i}(t) - \langle \vec{r_i}\rangle_{\delta t}]^2\rangle_{\delta t,\cal{C}}$, where the
average is over the isoconfiguration ensemble as well as over a short
time interval $\delta t$ which in this work is taken to be $25$.

Starting from the same equilibrated configuration, we have run $6$
isotrajectories and have obtained the cage jumps for all of
them. Remarkably, all of the cage jumps occurring in the interval of
time $[t,t+\tau_S]$ fall on top of high DW areas, see
fig.~\ref{fig:DW}-top. Note that $\tau_S\gg25$, thus the correlation
between the DW map at time $t$ (a nearly instantaneous structural
quantity), and the dynamics taking place at longer times, is
non-trivial.  

Comparing the two top panels of Fig~\ref{fig:DW}, we find that different
isoconfigurational trajectories lead to cage jumps that take place at
different times and in different regions although they are always
located on top of high DW areas. This means that although clusters are
very likely to be in soft regions, when and where they exactly appear
is a stochastic event.  The two top panels of Fig~\ref{fig:DW}
strongly imply that a significant part of the avalanche structure of facilitated motion,
and not just the initial cluster in an avalanche, occurs on top of the
real-space geometric structure encoded in the soft mode map. 
Remarkably, we find that merging all cage jumps that occur in the
interval of time $\tau_S$ in the $6$ isoconfigurational trajectories
cover nearly all the high DW areas, as shown in fig.~\ref{fig:DW}
(bottom left).  A similar comparison with localized low frequency normal
modes, along the lines of~\cite{harrowell2008irs}, shows less, but
still significant, correlation. We interpret this as a signature
of anharmonic effects appearing in the vibrational structure of our
model of a supercooled liquid. Indeed, it is likely that there are
several potential energy minima in the basin in which the liquid is
confined at short times. The local DWs allow one to overcome this
difficulty and still provide a measure of local softness.
The above results are in agreement with the previous
conclusion of Berthier and Jack~\cite{berthierjack}, who found that
structural properties are better predictors of dynamics on large as
opposed to short length scales.

In order to present a more quantitative proof of the correlation
between DWs and cage jumps, we have computed the $DW$ at time $t$
averaged only over particles that jump between $t$ and $t+\tau$ as a
function of the lag time $\tau$. This quantity, normalized with
respect to $\langle DW(t) \rangle_{t}$ for all particles, is shown in
fig.~\ref{fig:DW} (bottom right). We find that at short times the
average $DW$ for the jumping particles is substantially higher than
the $DW$ averaged over all particles. This correlation disappears for
larger times comparable to times over which the DW maps decorrelate,
which we find to be roughly of the order of $\tau_\alpha/3$.

\begin{figure}[t!] 
\center
\vspace{-0.5cm}
\includegraphics[height=0.47\columnwidth]{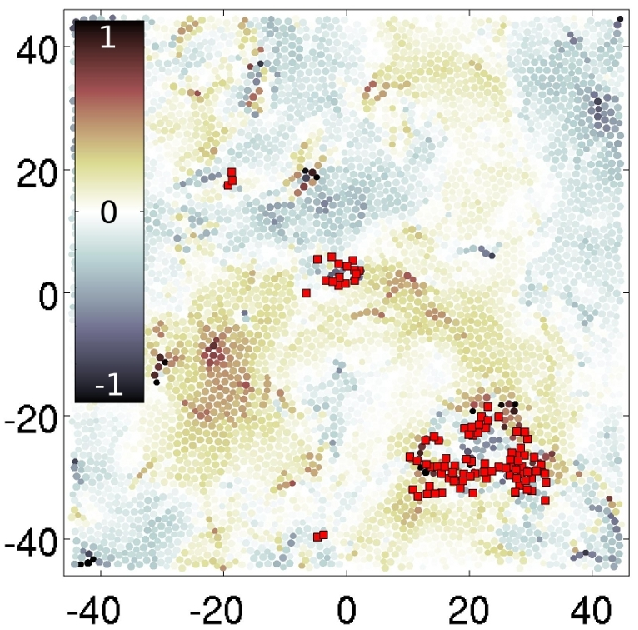}
\includegraphics[height=0.47\columnwidth]{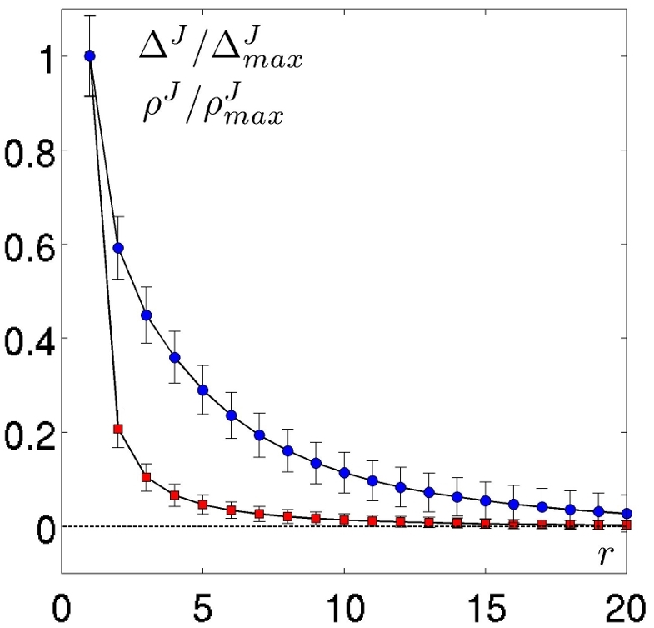}
\caption{(color online) Left: Cage jumps occurring in $\tau=17$ on top of a map of the relative difference $(DW(t+\tau)-DW(t))/\langle DW\rangle$.  Right: Normalized $\Delta^J(r) = \langle |\delta DW_i|\rangle^J_r - \langle |\delta DW_i|\rangle^J_\infty$ (blue circles) where $\langle |\delta DW_i| \rangle^J_r$ is the absolute difference of $DW$ over $\tau=17$ averaged over the particles in the disk of radius $r$ around a given cage jump. The analogous quantity for the density of jumps $\rho^J(r) = \langle \delta_i \rangle^J_r - \langle \delta_i \rangle^J_\infty$ (red squares) where $\delta_i$ is $1$ if particle $i$ jumps between $t$ and $t+\tau$ and $0$ otherwise. Error bars are given by the standard deviation.  }
\label{fig:corr}
\vspace{-0.5cm}
\end{figure}

A final issue worth investigating concerns the relation between cage jumps and DW map renewal.
We find that decorrelation is a distinctly non-local process. More precisely,
we have discovered that a cage jump at time $t$ correlates with changes of
the DWs that happen shortly after and extend quite far away.
This is demonstrated visually in fig.~\ref{fig:corr}(left).
In order to provide a quantitative proof we consider $|DW(t)-DW(t+\tau)|$
averaged over all particles, that are at distance $r$ from a cage
jump taking place at time $t$ and subtract from that quantity its
$r=\infty$ value. In fig. \ref{fig:corr}(right) we show this quantity,
called $\Delta^J(r)$, for $\tau=17$.  One finds that $\Delta^J(r)$ is
quite long ranged, in particular much more than the cage jump
correlation function $\rho^J(r)$, see fig. \ref{fig:corr}(right) and
its caption for a precise definition of $\rho^J(r)$.  What is
mediating the non-local interaction between cage jumps and DWs is an
intriguing question.  One possibility is that a slowly varying spatial
field, like the thermal strain discussed in~\cite{harrowellstrain},
plays an important role by providing long ranged dynamical
interactions.

The picture that emerges from our study is that the dynamics occurs,
as in the dense granular system studied in~\cite{candelier2009bbo,candelier2009eod},
via a two time scale process that gives rise to dynamical
heterogeneities and induces macroscopic relaxation. At short times,
the particles collectively jump within clusters whose sizes are very
widely distributed. These clustered jumps trigger other ones nearby,
leading to well separated large scale avalanches.

We find that this dynamical facilitation process is coupled 
to the structure : mobility preferentially follows the soft 
regions and has a non-local influence on the evolution of the
topography of hard and soft areas. The resulting picture of
facilitation is quite different from the one based on the
propagation of a conserved mobility field.

Studying the evolution of dynamical properties with decreasing
temperature following the same analysis would allow for direct
tests of prominent theories of the glass transition.
For example, in the picture based on kinetically constrained models of glasses
\cite{garrahan2002prl} facilitation should become more relevant and
conserved upon lowering the temperature. In the random first order
transition theory~\cite{RFOT}, the dynamics should be correlated
with soft regions for moderately supercooled liquids but,
closer to the glass transition, the relaxation should be
dominated by other processes. Three of us~\cite{candelier2009eod}
have performed such analysis for granular media and found that
facilitation becomes {\em less} conserved as the density is increased.
Performing a similar analysis for our model of supercooled liquids
would be extremely important. Work in this direction is in progress.

We would like to thank J.-P. Bouchaud and L. Berthier for fruitful discussions.
GB, RC and OD were partially supported by ANR DYNHET 07-BLAN-0157-01.  DRR
would like to thank the National Science Foundation for financial
support. PH is supported through the Discovery program of the Australian
Research Council. AW thanks the School of Chemistry at the University
of Sydney for computer time on the Silica cluster.

\vspace{-5mm}
\bibliography{biblio_glass}

\end{document}